\documentclass[english,aps,pre,a4paper,twocolumn,amsfonts,amsmath,amssymb,nofootinbib,10pt]{revtex4-1}
\usepackage[english]{babel}
\usepackage{graphicx}
\usepackage{epstopdf}
\usepackage{mathrsfs}
\usepackage{enumerate}
\usepackage{color}
\usepackage{hyperref}

\begin{document}

\def\Journal#1#2#3#4#5#6#7{{#1}, {\it #4} \textbf{#5}, #6 (#2).}
\def\Book#1#2#3#4#5{{#1}, {\it #3} (#4, #5, #2).}

\newcommand{\dd}{\mbox{d}}
\newcommand{\NN}{\mathbb{N}}
\newcommand{\ZZ}{\mathbb{Z}}
\newcommand{\RR}{\mathbb{R}}
\newcommand{\PP}{\mathbb{P}}
\newcommand{\EE}{\mathbb{E}}
\newcommand{\uu}{\mathbf{1}}
\newcommand{\RM}{\mathscr{R}}
\newcommand{\FM}{\mathscr{F}}
\newcommand{\NM}{\mathscr{N}}
\newcommand{\nn}{n}

\title{Sisyphus random walk}
\author{Miquel Montero}
\email[E-mail: ]{miquel.montero@ub.edu}
\affiliation{Secci\'o de F\'{\i}sica Estad\'{\i}stica i Interdisciplin\`aria, Departament de F\'{\i}sica de la Mat\`eria Condensada, Universitat de Barcelona (UB), Mart\'{\i} i Franqu\`es 1, E-08028 Barcelona, Spain}
\affiliation{Universitat de Barcelona Institute of Complex Systems (UBICS), Universitat de Barcelona, Barcelona, Spain}
\author{Javier  Villarroel}
\email[E-mail: ]{javier@usal.es}
\affiliation{Facultad de Ciencias \& Instituto Universitario de F\'{\i}sica Fundamental y Matem\'aticas, Universidad de Salamanca, Plaza Merced s/n, E-37008 Salamanca, Spain}
\date{\today}
\begin{abstract}
In this paper we consider a particular version of the random walk with restarts: random reset events which bring suddenly the system to the starting value.  We analyze its relevant statistical properties like  the transition probability and show how an equilibrium state  appears. Formulas for the first-passage time, high-water marks and other extreme statistics are also derived: we consider counting problems associated naturally to the system. Finally we indicate feasible generalizations useful for interpreting different physical  effects.
\end{abstract}
\maketitle

\section{Introduction}

The {\it Sisyphus random walk\/}, the stochastic process to be introduced in this paper, is an infinite Markov chain whose dynamics can be expressed as follows: at every clock tick the process can move rightward (or upward) one step, with a certain given probability, or else return to the initial state, from where it is restarted. Such  apparent  simplicity is misleading as this simple   evolution law can exhibit a surprisingly complex and rich behavior

The process considered here bears some analogy with the punishment, in the Greek mythology, of Sisyphus, the first king of Ephyra, who was sentenced to climb up a hill carrying a heavy, slippery  boulder and  watch it roll down again to the starting point in an endless cycle. Remarkably, it also underlies the behavior of some physical mechanisms. In the context of Doppler laser cooling, ``Sisyphus effect'' is a well known mechanism by which alkali atoms in the presence of a light field raise from the ground Zeeman level to higher excited states or sub-levels. This {\it up hill\/} climbing process increases the probability to be optically pumped into a minimum potential-energy state from where the process restarts.  In addition, it  involves a loss of momentum so after each Sisyphus cycle, the total energy of the atom decreases by a certain amount.~\footnote{In 1997 the Nobel Prize in Physics  was awarded to Steven Chu, William Phillips and Claude Cohen-Tannoudji  ``for development of methods to cool and trap atoms with laser light,'' see the lectures of the three Nobel Laureates~\cite{NL97}.} 

In a different setting, such system may be used as an idealized model of the random dynamics of a ``mobile" in a trap, say, who is trying to climb stepwise a ladder or wall given that at every step there is a common probability  of slipping to the bottom, resulting in the need to restart again. In such situation the distribution of the time to escape the trap becomes a natural question.

The hallmark of such processes is the possibility to display return-to-the-origin behavior, a common feature in real life systems. The seminal work of Manrubia and Zanette~\cite{MZ99} considering Markov chains where a reset mechanism operates has motivated new interest in the field and presently the dynamics of systems with resets is being subjected to intense study. In~\cite{EM11a,EM11b} Brownian motion with resets was considered while in~\cite{MV13}  resets were incorporated to a compound Poisson process with constant drift. Such intermittent strategies have been considered in general mathematical frameworks~\cite{EM11a,EM11b,MV13,JP12,EM13a,EM14}, but also in more specific contexts, like behavioral ecology where the browsing activity of living organisms (as, e.g., capuchin monkeys) may be suddenly interrupted to return to a preferred location~\cite{BSS,KMSS}, or econophysics where, modifying Gibrat's law to include reset events, has made it possible to account for the power law's distribution of firm's growth~\cite{Aoyama}. The present paper continues this line of research and considers a different special case of stochastic processes with a reset mechanism. See~\cite{GMS,Durang,MSS,BP16,MC16} for other developments in this regard.

The paper is organized as follows. In Sec.~\ref{Sec_process} we introduce the process at hand and remind some basic concepts of renewal theory. In Sec.~\ref{Sec_transition}  we derive an explicit expression for the  {\it propagator\/}, the transition probability function of the process, an essential magnitude for understanding the dynamics that provides a paradigmatic example of the use of renewal concepts. The statistics analysis  of several extreme events as, e.g., the first-passage time or the maximum of the process, and related  survival probabilities, is discussed  in Secs.~\ref{Sec_extreme} and~\ref{Sec_water_mark} where we also analyze in detail related counting problems and recurrence issues. In Sec.~\ref{Sec_alternative} we generalize the model to include different physical effects. By allowing the reset probability to be {\it  random\/}  or {\it site-dependent\/} interesting generalizations are obtained. The resulting random walk may be used to model physical systems which  become ``increasingly anxious'' to restart as they drift far away from the initial state, as happens in situations as diverse as Sisyphus cooling or formation |and eventual collapse| of  built-up structures like stalagmites. Depending on the election for the reset probability we find that the equilibrium state has heavy or light tails and obeys a geometric, a zeta or a Poisson distribution. Other possible generalizations include the interesting  possibility of having a random walk on the  integers which upon reset   may drift upward or downward with different probabilities. Here the problem of finding the optimal reset strategy appears naturally. Conclusions are drawn in Sec.~\ref{Sec_conclusions} and, finally, we complete some technical details in the appendix.
 
\section{Sisyphus random walk as a renewal process}
\label{Sec_process}

The Sisyphus random walk, $X_t$, is an infinite Markov chain on the positive integers |namely $X_t\in\{0,1,2,\ldots\}$ for $t~\in~\{0,1,2,\ldots\}$| whose one-step evolution can be expressed as follows: If  at time $t$ the walker is at a given location, $X_{t}=\ell$, then at time $t+1$ one has
\begin{equation}
X_{t+1}=\left\{
\begin{array}{ll}
\ell+1,&\mbox{ with probability } q_{\ell},\\
0,&\mbox{ with probability } (1-q_{\ell}),
\end{array}
\right.
\label{process}
\end{equation}
that is, at every clock tick the process can increase one unit, or return to the {\it ground state\/}, from where the evolution continues. If the initial condition is set in such a way that $X_{0}=0$, as we do consider, this return to the ground state can be understood as a restart of the process. 

\begin{figure}[htbp]
{
\includegraphics[width=0.9\columnwidth,keepaspectratio=true]{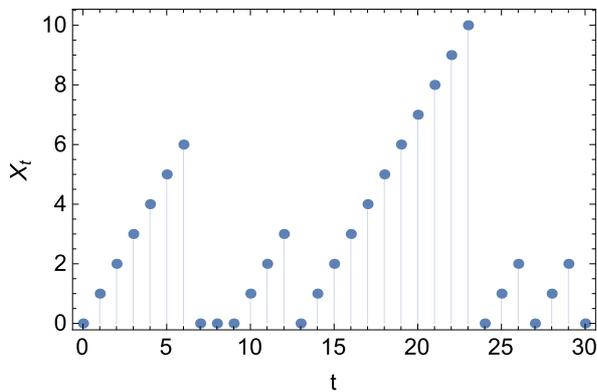}
}
\caption{
Typical realization of the process $X_t$ for $q_{\ell}=4/5$, independent of $\ell$.} 
\label{Fig_sample}
\end{figure}

Typical sample paths of the process $X_t$ are (irregular) sawtooth functions, namely, piecewise linear functions that slope upward and then sharply fall at reset times, see Fig.~\ref{Fig_sample}.

Equation~\eqref{process} stresses the fact that the elements $q_{\ell}$ of the transition matrix between different locations on the chain can depend on the present state $\ell$ of the system, an assumption consistent with the Markov property. We rule out, however, the possibility that the transition matrix depends on the chronological time $t$, or on some other hidden variable. Under such hypothesis, the process is time homogeneous. 

An appropriate dependence of $q_{\ell}$ on $\ell$ may befit a model  where, as a result of learning abilities |or some exogenous circumstance| the system becomes more (or less) anxious to return to the origin the farther off it is. In the rest of the paper we study the most relevant statistical magnitudes of $X_t$ corresponding to the case $q_{\ell}=q$, a constant parameter. This simplification reduces the algebraic intricacy of a model which, despite this, is still mathematically rich and complex. We indicate some results  corresponding to different $\ell$-dependence in Sec.~\ref{Sec_alternative}.   

To this end we note that, under such assumption, the existence of resets gives rise to an underlying renewal structure~\footnote{Discrete renewal theory, as used here, is a prototype tool in fields like reliability theory or block replacement policies, see~\cite{kt}.} and hence, for most statistics of interest, renewal-type equations can be employed to advantage over conventional random walk theory, see~\cite{klafter}. The reader is also referred to the interesting paper~\cite{KMSS}, where some properties of a related discrete-time random walk |which evolves via {\it L\'evy flights\/} on the line combined with resets| are considered.

A key quantity in renewal theory is the renewal function, $m(t;t_0)$, the mean number of reset events in a given interval $(t_0,t]$. Most of the formulas employed in the paper consider different properties of the walker in some future instant $t$, subject to the knowledge of its state at time $t_0$, $t_0<t$. Due to the time homogeneity of the process, those expressions actually depend not on the calendar time but on the time lapsed between the two events, $\tau\equiv t-t_0$, and thus we define $m(\tau)\equiv m(t_0+\tau;t_0)$. 

The renewal function $m(\tau)$ satisfies a renewal equation, which reflects the two possible scenarios that appear depending on the relative value of $\tau$ with respect to $\tau^*$, the random instant at which the {\it first\/} reset (after $t_0$) takes place. Using that
 \begin{equation}
\PP\left\{\tau^*=k\right\}=q^{k-1} (1-q) \text{ and } \PP\left\{\tau^*>\tau\right\}=q^{\tau},
\label{reset_time}
\end{equation} where  $\PP\{\cdots\}$ is   the probability relative  to the set $\{\cdots\}$ one finds that $m(\tau)$ must solve  
 \begin{eqnarray}
m(\tau)&=& 1-q^{\tau}+\sum_{k=1}^{\tau} q^{k-1} (1-q) m(\tau-k).
\label{renew_m}
\end{eqnarray}
Equations like~\eqref{renew_m}  can be conveniently analyzed by recourse to the so called $z$-transform:  
\begin{equation*}
\widehat{m}(z)\equiv\sum_{k=0}^{\infty}m(k)z^k,
\end{equation*}
where $z$ is a complex variable. Equation~\eqref{renew_m} is solved in terms of this object to find 
\begin{equation}
\widehat{m}(z)=\frac{(1-q) z}{(1-z)^2}, \text{ and }  
m(\tau)=(1-q)\tau=\frac{\tau}{\EE\left[\tau^*\right]},
\label{m_sol}
\end{equation} 
where $\EE[\cdot]$ denotes the expectation of its argument. 

\section{Transition probability}
\label{Sec_transition}

\begin{figure}[htbp]
{
\includegraphics[width=0.9\columnwidth,keepaspectratio=true]{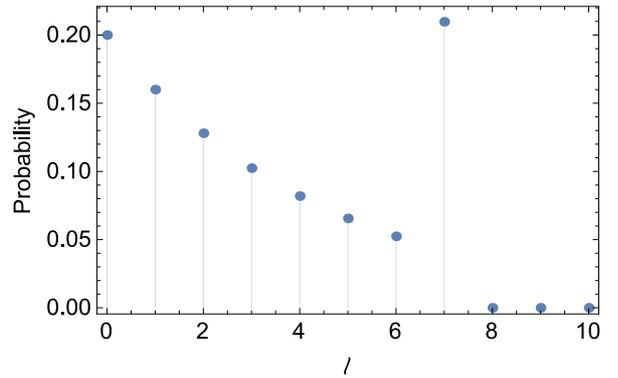}
}
\caption{
Distribution of the process $X_t$ at time $t=7$ for $q=4/5$. Note that the probability is decreasing, but the most likely position is found at the ending point $\ell=t$.} 
\label{Fig_dist}
\end{figure}

We start our analysis of $X_t$ with the determination of the transition probability of the process, the {\it propagator\/}:
\begin{equation}
p(\ell,t;\ell_0,t_0)\equiv \PP\left\{\left. X_t =\ell\right| X_{t_0}=\ell_0\right\}.
\end{equation}
The propagator gives the probability of finding the walker in position $\ell$ at the future instant $t$, if we known that it is in $\ell_0$ at the present time $t_0<t$. Note that, as discussed previously, it depends on time only via $\tau=t-t_0$ but must depend on both $\ell$ and $\ell_0$:
\begin{equation*}
p(\ell,t;\ell_0,t_0)=p(\ell,t-t_0;\ell_0,0)\equiv p(\ell,\tau;\ell_0).
\end{equation*}
Indeed we do not expect translation invariance since the reset mechanism favours a particular point, the origin; as a consequence the probability function for the walker position
\begin{equation*}
p(\ell,t;0)\equiv \PP\left\{\left. X_t =\ell\right| X_{0}=0\right\}
\end{equation*}
does not contain all physical information and {\it the full transition probability\/},  Eq.~\eqref{TP_sol} below, {\it is required to describe the dynamics\/}.  
The  equation that  $p(\ell,\tau;\ell_0)$ satisfies follows by analyzing the different situations that may present depending on whether $t_0+\tau^*$, namely the first reset  after $t_0$, occurs  or not in the interval $(t_0,t_0+\tau]$. If  $\delta_{k,k'}$ is the Kronecker delta, the transition probability becomes  $p(\ell,\tau-\tau^*;0)$  in the first case and $\delta_{\tau,\ell-\ell_0}$  in the latter |see also~\cite{MSS} for a similar reasoning in the context of Brownian motion with resets to the origin.  In view of all this, and the probabilities~\eqref{reset_time}, $p(\ell,\tau;\ell_0)$ must satisfy the following equation:
\begin{eqnarray}
p(\ell,\tau;\ell_0)&=&q^{\tau} \delta_{\tau,\ell-\ell_0}
+\sum_{k=1}^{\tau}q^{k-1} (1-q)  p(\ell,\tau-k;0).\nonumber\\
\label{TP}
\end{eqnarray}
Letting first  $\ell_0=0$ in~\eqref{TP} an equation for $ p(\ell,\tau;0)$ follows which can be handled by using $z$-transforms. We find 
\begin{equation}
p(\ell,\tau;0)=q^{\tau}\delta_{\tau,\ell} +(1-q)q^{\ell} \Theta\left(\tau-\ell-1\right),
\label{TP0_sol}
\end{equation}
where $\Theta\left(u\right)$ is the (right-continuous) Heaviside step function: $\Theta\left(u\right)=1$ for $u\geq0$, and $\Theta\left(u\right)=0$ for $u<0$. Hence, having started from the origin the position has  (truncated) geometric distribution $p(\ell,\tau;0)$, which decreases with  $\ell$. However,  if $q>1/2$, the mode of the distribution is found at $\ell=\tau$, as Fig.~\ref{Fig_dist} shows.

By insertion of~\eqref{TP0_sol} in Eq.~\eqref{TP} we obtain the complete propagator as
\begin{equation}
p(\ell,\tau;\ell_0)=q^{\tau}\delta_{\tau,\ell-\ell_0} +(1-q)q^{\ell} \Theta\left(\tau-\ell-1\right).
\label{TP_sol}
\end{equation}
Hence the only effect of having started from $\ell>0$ is to {\it shift the highest accesible site\/} from $\tau$ to $\tau+\ell_0$, maintaining the probabilities of the remaining accessible states $\{0,1,\dots,\tau-1\}$ unchanged:
\begin{equation} 
p(\ell,\tau;\ell_0)=p(\ell,\tau;0),\  0\le \ell \le\tau-1, 
\label{semisymm}    
\end{equation}
and $p(\ell_0+ \tau,\tau;\ell_0)=p(\tau,\tau;0)$.

\begin{figure}[htbp]
{
\includegraphics[width=0.9\columnwidth,keepaspectratio=true]{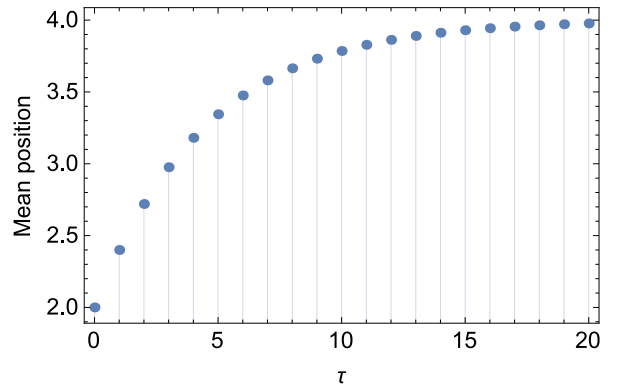}
}
\caption{
The conditional mean position of the process, $\EE\left[\left.X_{t}\right|X_{t_0}=\ell_0\right]$, is plotted as a function of $\tau=t-t_0$, for $\ell_0=2$ and $q=4/5$.} 
\label{Fig_mean}
\end{figure}

From  Eq.~\eqref{TP_sol}  the conditional mean position is given by
\begin{equation}
\EE\left[\left.X_{t}\right|X_{t_0}=\ell_0\right]=\sum_{\ell=0}^{\infty}\ell\, p(\ell,\tau;\ell_0)=\frac{q-q^{\tau+1}}{1-q}+q^{\tau} \ell_0,
\label{mean_pos}
\end{equation}
and is a monotone function of $\tau$ going from $\ell_0$ to the limit value $q/(1-q)$, see Fig.~\ref{Fig_mean}. 

Figures~\ref{Fig_sample} to~\ref{Fig_mean} show the sample path, distribution and, respectively, the conditional mean of the processes $X_t$ when $q=4/5$ (and $\ell_0=2$). In this case the  average time between resets is $5$ and the average limit value is $4$ |see Eqs.~\eqref{reset_time} and~\eqref{mean_pos} above. Notice how the mean function fails to capture the sharp drops in trajectories at reset times.

From Eq.~\eqref{TP_sol} we obtain the stationary state
\begin{equation}
p(\ell)\equiv\lim_{\tau\rightarrow \infty} p(\ell,\tau;\ell_0)=\lim_{\tau\rightarrow \infty} p(\ell,\tau;0)= (1-q)q^{\ell},
\label{stat_PDF}
\end{equation}
a geometric distribution with mean $\mu=q/(1-q) $ and typical standard deviation $\sigma=\sqrt{q}/(1-q)$. Thus the system is recurrent and ergodic and, given enough time, attains an equilibrium distribution. This ergodicity could be expected on physical grounds since the incorporation of this reset mechanism guarantees that the system will not be driven too far off from the origin. Note that this stationary state $X_\infty \equiv \lim_{\tau\rightarrow \infty} X_\tau$  has the same distribution as $X_{\tau^*-1}$, the distance covered before the motion restarts: 
\begin{equation}
\PP\left\{X_{\tau^*-1}=\ell\right\}= \PP\left\{X_{\infty}=\ell\right\},
\label{match}
\end{equation}
cf. Eqs.~\eqref{reset_time} and~\eqref{stat_PDF}. We stress that those remarkable concurrences in Eqs.~\eqref{semisymm}  and~\eqref{match} do not extend to general election of $q_\ell$. 

\section{Extreme-time statistics}
\label{Sec_extreme}

In the next two sections we study some statistical properties of several {\it extreme\/} functionals associated to the process $X_t$. In our context, an extreme event is simply a physical observable or quantity related to $X_t$ that has attained a minimum or a maximum. It turns out that there are several of them which are quite interesting from a physical viewpoint.

\subsection{First-passage statistics}
\label{Sec_first}

Let $\ell$ be a given level. The first-passage time, or hitting time~\cite{JP12}, $\mathcal{F}({\ell})$,
\begin{equation}
\mathcal{F}({\ell})\equiv\min \left\{t: X_t=\ell | X_0=0\right\},
\label{def_F}
\end{equation}
represents the minimum lapse of time needed for the process to travel from the ground state to the given site, or for the walker to exit the trap by climbing an elevation $\ell$.  We begin considering the analysis of  the first-passage time probability of the process to this level, $\mathcal{P}(t,\ell)$,
\begin{eqnarray}
\mathcal{P}(t,\ell)
&=& \PP \left\{\mathcal{F}({\ell})=t \right\},
\label{FP_def}
\end{eqnarray}
namely, the probability that the process which is initially at $X_0=0$ reaches level $\ell$ for the first time at instant $t$.


The renewal equation for $\mathcal{P}(\ell,t)$ reads
\begin{eqnarray}
\mathcal{P}(t,\ell)&=&q^{\ell} \delta_{t,\ell}+\sum_{k=1}^{\ell} q^{k-1} (1-q)  \mathcal{P}(t-k,\ell),
\label{FP_renew}
\end{eqnarray}
where the first term accounts for the eventuality that the process reaches $\ell$ without restarting, and the summation contains those cases in which the first reset takes place at time $t^*=k\leq \ell$. (We remark that related techniques have been used elsewhere in the context of reset systems; note, in particular, the close similarity with the derivation of the survival probability 
in~\cite{KMSS}.)

\begin{figure}[htbp]
{
\includegraphics[width=0.9\columnwidth,keepaspectratio=true]{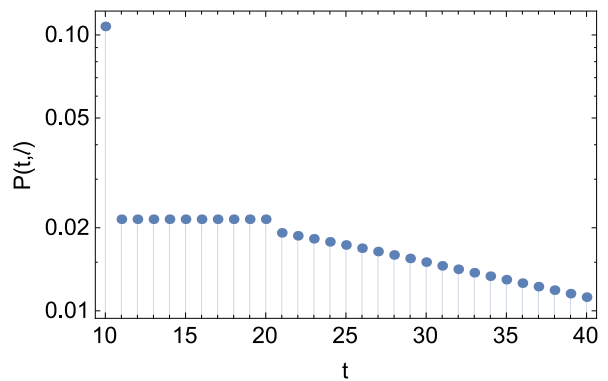}
}
\caption{
First-passage time probability of the processes $X_t$, for $\ell=10$ and $q=4/5$.} 
\label{Fig_time}
\end{figure}

Taking the $z$-transform of~\eqref{FP_renew} with respect to the time variable $t$ one gets
\begin{equation}
\widehat{\mathcal{P}}(z,\ell)=\left(q z\right)^{\ell}\frac{1-q z}{1- z+ (1-q)z\left(q z\right)^{\ell}}.
\label{zFP_sol}
\end{equation}
The value of any given $\mathcal{P}(t,\ell)$ could be obtained from the 
$t$-order derivative of $\widehat{\mathcal{P}}(z,\ell)$ at $z=0$,
\begin{equation*}
\mathcal{P}(t,\ell)=\left.\frac{1}{t!}\frac{\partial^t \widehat{\mathcal{P}}(z,\ell)}{\partial z^t}\right|_{z=0},
\end{equation*}
and leads to the following formula, valid for $t\geq \ell+1$, see App.~\ref{zTransform}:
\begin{eqnarray}
\mathcal{P}(t,\ell)&=&q^{\ell} \sum_{k=0}^{\left\lfloor\frac{t-\ell}{\ell+1}\right\rfloor} \binom{t-(k+1)\ell}{k}\left[(q-1)q^{\ell}\right]^k\nonumber \\
&-&q^{\ell+1} \sum_{k=0}^{\left\lfloor\frac{t-\ell-1}{\ell+1}\right\rfloor}\binom{t-(k+1)\ell-1}{k}\left[(q-1)q^{\ell}\right]^k, \nonumber \\
\label{Pt_sol}
\end{eqnarray}
where $\lfloor \cdot \rfloor$ denotes the {\it floor function\/},
\begin{equation*}
\lfloor x \rfloor=\max\, \{k\in\mathbb{Z}\mid k\le x\}.
\end{equation*}
In Fig.~\ref{Fig_time}, we plot the corresponding distribution probability for $\ell=10$ as a function of $t$.

It is fortunate that despite the unwieldiness of expression Eq.~\eqref{Pt_sol} most statistical magnitudes can be obtained in an easy way. In particular, the mean first-passage time 
can be easily obtained from $\widehat{\mathcal{P}}(z,\ell)$: 
\begin{eqnarray}
\EE\left[\mathcal{F}(\ell)\right]=\left.\frac{\partial \widehat{\mathcal{P}}(z,\ell)}{\partial z}\right|_{z=1}=\frac{1}{1-q}\left(\frac{1}{q^{\ell}}-1\right),\label{meanhit}
\end{eqnarray}
and hence  the mean  time to reach $\ell$ grows exponentially with the distance~\cite{EM11a}.

\subsection{Record statistics}
\label{Sec_record}

By construction, the first time our Sisyphus random walk visits a given threshold $\ell$, $\mathcal{F}({\ell})$, it scores a {\it record\/}: the highest value reached by the process up to that time. Since the process is right-continuous, one can guarantee that  $X_{t'}<\ell$ for $0\leq t'<\mathcal{F}({\ell})$, and $X_t=\ell$, at $t=\mathcal{F}({\ell})$.

The next question that naturally arises within this context is what are the properties of the interval spent between two records of the process~\cite{Godre}, the inter-record statistics? To answer this question we are going to generalize~\eqref{def_F} and define $\mathcal{F}(\ell;t_0,\ell_0)$, 
\begin{equation}
\mathcal{F}(\ell;t_0,\ell_0)\equiv\min\left\{t> t_0:\left. X_t=\ell\right|X_{t_0}=\ell_0\right\},
\end{equation}
the first time the process reaches $\ell$ {\it after\/} time $t_0$, and call $\mathcal{P}(t,\ell;t_0,\ell_0)$ its associated probability
\begin{eqnarray}
\mathcal{P}(t,\ell;t_0,\ell_0)
&\equiv&\PP\left\{\mathcal{F}(\ell;t_0,\ell_0)=t\right\}.
\end{eqnarray}
When $\ell>\ell_0$, as we assume along in this section, $\mathcal{P}(t,\ell;t_0,\ell_0)$ is the probability that at time $t$, $t>t_0$, the process achieves the {\it new local maximum\/} $\ell$, since $X_{t'}<\ell$, for $t_0\le t'<\mathcal{F}(\ell;t_0,\ell_0)$. We will clarify later on the connection between $\mathcal{P}(t,\ell;t_0,\ell_0)$ and $\mathcal{R}(t,\ell;t_0,\ell_0)$, the probability that the process sets a new record $\ell$ at time $t$, provided it scored record $\ell_0$ at time $t_0$.

As in Sec.~\ref{Sec_transition}, the time-homogeneity of the process implies that $\mathcal{P}(t,\ell;t_0,\ell_0)$ is a function of $t$ and $t_0$ through the time interval $\tau=t-t_0$, $\mathcal{P}(\tau,\ell;\ell_0)\equiv \mathcal{P}(t-t_0,\ell;0,\ell_0)$.
In this case, the equation for $\mathcal{P}(\tau,\ell;\ell_0)$ reads 
\begin{equation}
\mathcal{P}(\tau,\ell;\ell_0)=q^{\tau} \delta_{\tau,\ell-\ell_0}+\sum_{k=1}^{\ell-\ell_0}q^{k-1} (1-q)  \mathcal{P}(\tau-k,\ell),
\label{P_renew}
\end{equation}
with the first term representing the contingency in which the process increases steadily during $\tau$ consecutive steps, passing from $\ell_0$ to $\ell$, whereas the summation contains those cases where the first reset happens after a lapse of $\tau^*=k\leq \ell-\ell_0$. Since the restart takes the process to the origin, in the sum appears $\mathcal{P}(\tau,\ell)=\mathcal{P}(\tau,\ell;0)$, see Eq.~\eqref{FP_def}. 

The $z$-transform of Eq.~\eqref{P_renew} with respect to $\tau$ reads:
\begin{eqnarray}
\widehat{\mathcal{P}}(z,\ell;\ell_0)
&=&\frac{(1-z)\left(q z\right)^{\ell-\ell_0}+(1-q) z \left(q z\right)^{\ell}}{1- z+ (1-q)z\left(q z\right)^{\ell}},
\label{zP_sol}
\end{eqnarray}
where we have used Eq.~\eqref{zFP_sol}. The general expression of $\mathcal{P}(\tau,\ell;\ell_0)$ follows by differentiation around $z=0$. 

Now, assume that  $X_{t_0}=\ell_0$ {\it is\/} a record. If such is the case, $X_{t'}<\ell_0<\ell$, for $0\leq t'<t_0$. We also know that $X_{t'}<\ell$, for $t_0\leq t'<t$, consequently, $X_{t'}<\ell$, for  $0\leq t'<t$, and the process scores a new record at time $t$. In conclusion, the inter-record probability, $\mathcal{R}(\tau,\ell;\ell_0)$, is equal to $\mathcal{P}(\tau,\ell;\ell_0)$ for $\ell>\ell_0$.

Basic moments of $\mathcal{R}(\tau,\ell;\ell_0)$ follow from Eq.~\eqref{zP_sol} by taking derivatives at $z=1$. In particular, the mean inter-record lapse is
\begin{eqnarray}
\RM(\ell;\ell_0)
=\left.\frac{\partial \widehat{\mathcal{R}}(z,\ell;\ell_0)}{\partial z}\right|_{z=1}=\frac{1}{1-q}\left[\frac{1}{q^{\ell}}-\frac{1}{q^{\ell_0}}\right].
\end{eqnarray}

Finally, note that we have not demanded that the two records are {\it consecutive\/}, that is, that there is no additional record in the time interval. In our case, however, the statistics associated to this contingency can be easily derived from $\mathcal{R}(\tau,\ell;\ell_0)$, by merely setting $\ell=\ell_0+1$, $\mathcal{R}(\tau,\ell)\equiv\mathcal{R}(\tau,\ell;\ell-1)$. Thus, for instance,
\begin{eqnarray}
\RM(\ell)\equiv\RM(\ell;\ell-1)=\frac{1}{q^\ell}.
\end{eqnarray}

\subsection{Mean recurrence time} 

In this section we will finish the analysis of $\mathcal{P}(\tau,\ell;\ell_0)$ when  $\ell \leq \ell_0$.
In this case, the system can only attain $\ell$ having first been reset to the origin, and therefore
\begin{equation}
\mathcal{P}(\tau,\ell;\ell_0)=\sum_{k=1}^{\infty}q^{k-1} (1-q)  \mathcal{P}(\tau-k,\ell),
\end{equation}
whose generating function is given by
\begin{eqnarray}
\widehat{\mathcal{P}}(z,\ell;\ell_0)
&=&\frac{(1-q) z \left(q z\right)^{\ell}}{1- z+ (1-q)z\left(q z\right)^{\ell}},
\label{zRT_sol}
\end{eqnarray}
where we have used again~\eqref{zFP_sol}. 

Equation~\eqref{zRT_sol} can be inverted by the methods described above. 
In particular, as expected on intuitive grounds, the probability that the chain ever visits $\ell$, starting at $\ell_0\ge \ell $, is $\widehat{\mathcal{P}}(z=1,\ell;\ell_0)=1$,
and all states are positive recurrent. Actually, site  $\ell$ is visited infinitely often with probability $1$.

The  {\it mean recurrence time\/} satisfies  
\begin{equation} 
\EE\left[\mathcal{T}_{\ell\to\ell}\right] =\left.\frac{\partial \widehat{\mathcal{P}}(z,\ell;\ell)}{\partial z}\right|_{z=1}= \frac{1}{p(\ell)},
\end{equation}
as the classical ergodic theorem predicts~\cite{klafter}. Further, the {\it time average\/} of the process is given by 
\begin{equation*}  
\lim_{k\to\infty} \frac{1}{k+1}\sum_{t=0}^k X_t= \EE\left[X_\infty\right] =\frac{q}{(1-q)}. 
\end{equation*} 

\section{Water marks}
\label{Sec_water_mark}

\subsection{Number of sites visited}
\label{Sec_sites}

Another interesting magnitude, closely connected to $\mathcal{P}(t,\ell)$, is the high-water mark: the highest value $\ell$ that the process has reached for a {\it fixed time\/} $t$, namely
\begin{equation}
M_t\equiv \max\{\left.X_0, \ldots, X_t\right| X_0=0\}.
\end{equation}

\begin{figure}[htbp]
{
\includegraphics[width=0.9\columnwidth,keepaspectratio=true]{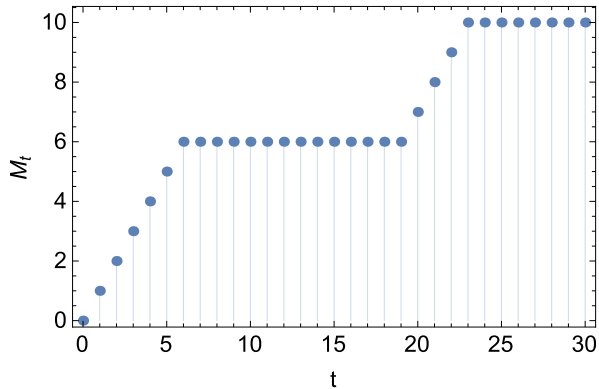}
}
\caption{
Realization of the process $M_t$ for the sample path $X_t$ shown in Fig.~\ref{Fig_sample}.} 
\label{Fig_sample_M}
\end{figure}

This quantity marks the threshold between those sites that have been already reached, from those that have not. Classical extreme-value theory is devoted to study the distribution of this statistics, typically assuming strong conditions on the increments of the process, i.e., independence and identically distributed, conditions that are not met here. Note that $0\le M_t\le t$ and, in contrast to $X_t$, the path of $M_t$ either increases linearly or remains constant |see Figs.~\ref{Fig_sample} and~\ref{Fig_sample_M}.

\begin{figure}[htbp]
{
\begin{tabular}{lc}
(a) & \includegraphics[width=0.9\columnwidth,keepaspectratio=true]{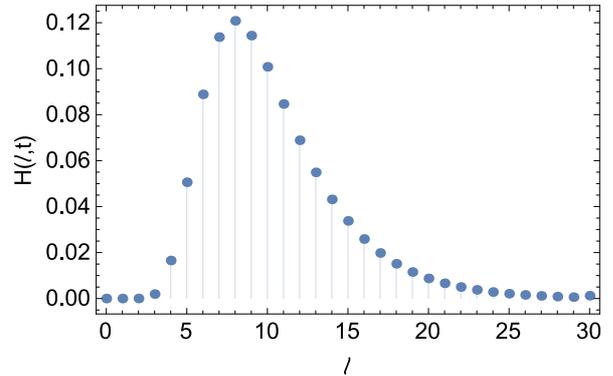}\\
(b) & \includegraphics[width=0.9\columnwidth,keepaspectratio=true]{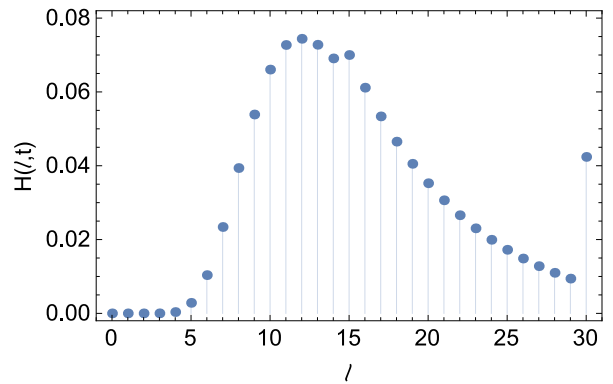}
\end{tabular}
}
\caption{
High-watermark probability of the processes $X_t$, for $t=30$ and (a) $q=4/5$; (b) $q=9/10$.} 
\label{Fig_high}
\end{figure}

Consider the probability associated to $M_t$, $\mathcal{H}(\ell,t)$,
\begin{eqnarray}
\mathcal{H}(\ell,t)&\equiv&
\PP\left\{M_t=\ell \right\}.
\end{eqnarray}
Obviously,  $\mathcal{H}(t,t)=q^t$ while  $\mathcal{H}(t-1,t)=2 q^{t-1}(1-q)$ since $M_t=t-1$ can only happen if the process suffers just one restart either at $t=1$ or at $t-1$. 
For general values, $0\le \ell \le t-2$, the derivation of the corresponding probabilities is not straightforward. The key fact is that  $M_t< \ell \Leftrightarrow \mathcal F(\ell) > t$. Hence  
\begin{equation}
\PP\left\{M_t< \ell \right\}= \PP\left\{\mathcal F(\ell) > t \right\}=\sum_{k=t+1}^{\infty}\mathcal{P}(k,\ell),
\label{M_t_cdf} 
\end{equation}
and therefore it follows
\begin{eqnarray} 
\mathcal{H}(\ell,t)&=&
\sum_{k=0}^{t}\left[\mathcal{P}(k,\ell)-\mathcal{P}(k,\ell+1) \right].
\end{eqnarray}
We can see a practical example in Fig.~\ref{Fig_high}, where for $t=30$ and two different choices for $q$, $q=4/5$ and $q=9/10$, we plot $\mathcal{H}(\ell,t)$ versus $\ell$. 
In the lower panel, observe the local maximum at $\ell=t$, and the kink at $\ell=t/2$. Exact determination of the mode of the distribution is not an easy task.

\subsection{Number of visits}
\label{Sec_visit}

Given a fixed time $t$, the high-water mark level counts the number of sites that have been visited. A related counting physical observable involves $N(t;\ell)$, the number of visits to a fixed level $\ell$ up to time $t$, 
and  $\mathcal{N}(\nn,t;\ell)$, the probability that this number equals $\nn\in \NN$, i.e.,
\begin{eqnarray}
\mathcal{N}(\nn,t;\ell)\equiv\PP\left\{N(t;\ell)=\nn \right\}.
\end{eqnarray}
In the extreme case $t=\ell$, obviously,  
\begin{eqnarray*}
\PP\left\{N(\ell;\ell)=1\right\}=1-\PP\left\{N(\ell;\ell)=0\right\}=q^\ell.  
\end{eqnarray*}
By contrast, if $\ell =0$, $N(t;0)$ is just  the number of resets up to time $t$ plus one (the initial visit) and hence $\EE\left[N(t;0)\right]=1+m(t)$. Finally the ergodic theorem guarantees that 
\begin{equation*}
\lim_{t\to\infty} \frac{N(t;\ell)}{t}= p(\ell),
\end{equation*}
with probability 1. In addition, the probability that the site $\ell$ have never been visited is just 
\begin{eqnarray}
\mathcal{N}(0,t;\ell)=\PP\left\{M_t<\ell \right\}.
\label{N_zero}
\end{eqnarray}

However, to go beyond these general statements and determine $\mathcal{N}(\nn,t;\ell)$ with all generality is far from trivial. We resort again to renewal arguments to obtain a set of equations with hierarchical structure:~\footnote{With no loss of generality we take  $\nn\le t$ and $\ell \le t$ since, at most, only a visit is possible within each reset interval and hence  $\mathcal{N}(\nn,t;\ell)=0$ if $\nn>t$ or $\ell >t$.} 
\begin{eqnarray}
\mathcal{N}(\nn,t;\ell)&=&q^{t} \delta_{\nn,1}
+\sum_{k=1}^{\ell}q^{k-1} (1-q)\,  \mathcal{N}(\nn,t-k;\ell)\nonumber\\
&+&\sum_{k=\ell+1}^{t}q^{k-1} (1-q)  \mathcal{N}(\nn-1,t-k;\ell).
\label{N_renew}
\end{eqnarray}
In this formula we can identify three different kinds of contributions. The single term assumes that no reset has taken place up to time $t$, and the process visits just once each of the levels between 0 and $\ell$. The first summation contains those cases for which the first reset takes place {\it before\/} the walker visits the targeted level $\ell$. In the second summation we find those cases in which the process has reached or passed $\ell$ once by the time of the first restart. 

The $z$-transform of Eq.~\eqref{N_renew} with respect to the time variable $t$ leads to
\begin{equation}
\widehat{\mathcal{N}}(\nn,z;\ell)=\frac{\left[ \delta_{\nn,1}+(1-q) z\widehat{\mathcal{N}}(\nn-1,z;\ell)\right] \left(q z\right)^{\ell}}{1- z+ (1-q)z\left(q z\right)^{\ell}},
\end{equation}
for $\nn \geq 1$, while the value of $\widehat{\mathcal{N}}(0,z;\ell)$ can be readily obtained from Eqs.~\eqref{zFP_sol}, \eqref{M_t_cdf}, and~\eqref{N_zero},
\begin{eqnarray}
\widehat{\mathcal{N}}(0,z;\ell)&=&\frac{ 1-  \left(q z\right)^{\ell}}{1- z+ (1-q)z\left(q z\right)^{\ell}}.
\end{eqnarray}
Hence, for $\nn\ge 1$,  the $z$-transform is given in an explicit way as
\begin{eqnarray}
\widehat{\mathcal{N}}(\nn,z;\ell)= \frac{ (1-qz)(1-q)^{\nn-1}z^{\nn-1}\left(q z\right)^{\nn \ell}}{\left[ 1- z+ (1-q)z\left(q z\right)^{\ell}\right]^{\nn+1}}.
\end{eqnarray}

Fortunately, there is no need to invert this expression to obtain the mean and main moments of the distribution. Indeed, say
\begin{eqnarray}
\widehat{\NM}(z;\ell)\equiv  \sum_{\nn=0}^{\infty} \nn\, \widehat{\mathcal{N}}(\nn,z;\ell) = \frac{\left(1-q z\right)\left(q z\right)^{\ell}}{\left(1-z\right)^2}.
\end{eqnarray}
 The   ``mean occupation number'', i.e., the mean number of visits to level $\ell$ at time $t\ge \ell$, follows by inversion of this result as
\begin{eqnarray}
\NM(t;\ell)=q^{\ell}\left[ 1+(1-q) \left(t-\ell\right)\right],
\end{eqnarray}
and depends linearly on $t-\ell$. In particular, letting $\ell=t$, $\ell=0$, or $t\to\infty$, the results at the beginning of the section are recovered.

\section{Alternative dynamics}
\label{Sec_alternative}

In this section we relax some of the assumptions made in the core of the main text, to indicate the capabilities and possible generalizations of Sisyphus random walk.

\subsection{Random probabilities}
\label{Sec_random}

The first variation to the previous setup is obtained assuming  that $q_{\ell}$ is still independent of $\ell$, but is random, and hence not a fixed parameter. This situation corresponds to a   walker climbing a ladder whose slip probability does not change with the step but remains unknown, due to insufficient information on the walker idiosyncrasy. By replacing $q$ by $Q$, a random variable,  some results appropriate to this case follow  from the previous expressions, by taking expectations with respect to $Q$. Concretely, unconditional probabilities may be derived this way. However, conditioning gives information which may partially pin down the slip probability of the walker.

For instance, suppose that
\begin{equation}
\PP\left\{q<Q\leq q+\dd q\right\}=\alpha (1-q)^{\alpha-1} \dd q,
\end{equation}
$\alpha>0$, i.e., $1-Q$ has a Pareto distribution on the interval $[0,1]$. Then one has that
\begin{equation}
\PP\left\{\tau^*=k\right\}=\alpha \frac{\Gamma(\alpha+1) (k-1)!}{\Gamma(\alpha+k+1)},
\label{YS}
\end{equation}
where $\Gamma\left(x\right)$ is the  Gamma function. This means that inter-reset times follow a Zipf-Simon-Yule (or discretized Pareto) distribution, well known in certain areas of econophysics, like wealth distribution. Recall that Eq.~\eqref{YS} is used to model the frequency of words in languages, or the size of objects randomly chosen from certain types of populations, see~\cite{Scalas}, pp. 260 and following. 

The mean value of the inter-reset time now reads
\begin{equation}
\EE\left[\tau^*\right]=\frac{\alpha}{\alpha-1},
\end{equation}
and is finite if and only if $\alpha>1$. Note that this excludes the uniform distribution, $\alpha=1$. Mean first-passage times or mean inter-record times are always unbounded magnitudes for $\ell\geq 1$.

The same law governs the properties of marginal and equilibrium probabilities of the process, cf. Eqs.~\eqref{TP0_sol} and~\eqref{stat_PDF},
\begin{eqnarray*}
p(\ell,\tau;0)&=& \frac{\Gamma(\alpha+1) \ell! }{\Gamma(\alpha+\ell+1)} \delta_{\tau,\ell} +  \frac{\alpha \Gamma(\alpha+1)  \ell! }{\Gamma(\alpha+\ell+2)}    \Theta\left(\tau-\ell-1\right),\\
p(\ell)&=&\lim_{\tau\to\infty} p(\ell,\tau;0)=   \frac{ \alpha \Gamma(\alpha+1)   \ell! }{\Gamma(\alpha+\ell+2)}.
\end{eqnarray*}  
Thus, the equilibrium distribution $p(\ell)$ has heavy tails with Pareto exponent  $\alpha+1$.


\subsection{Shrinking probabilities}
\label{Sec_shrinking}

\begin{figure}[htbp]
{ 
\includegraphics[width=0.9\columnwidth,keepaspectratio=true]{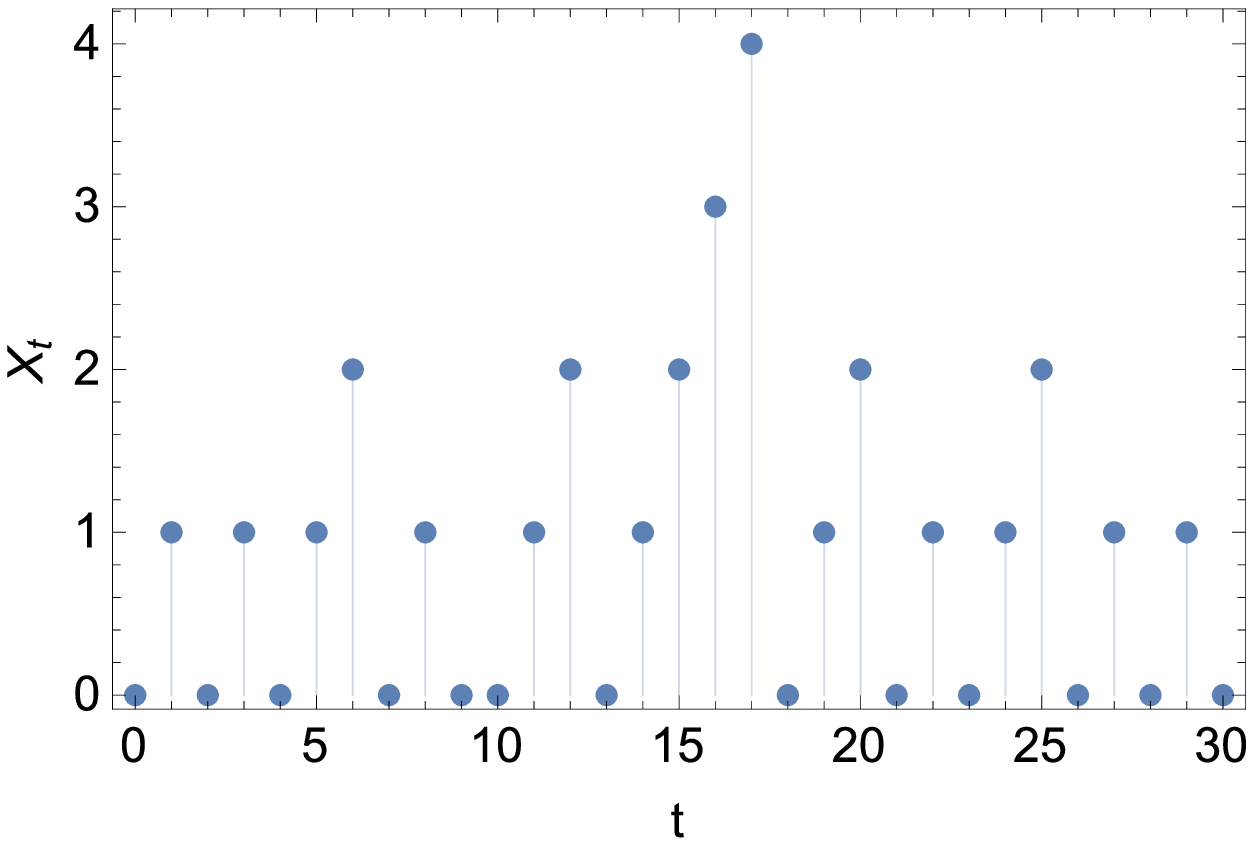}
}
\caption{
A feasible realization of the process $X_t$ with shrinking probabilities: $q_{\ell}=q_0/(\ell+1)$ and $q_0=4/5$.}
\label{Fig_shrinking}
\end{figure}

In this section we drop the requirement that $q_{\ell}$ be  constant~\cite{EM11b}. In this case most previous results, including the distribution of $\tau^* $  must be generalized appropriately. We focus on a system whose return probability increases with the location of the walker. This assumption is adequate to describe, say, the process of formation and eventual collapse of stalagmites or a   house of cards. A natural election is 
\begin{equation}
q_{\ell}=q_0 \left[1-\left(\frac{\ell}{\ell+1}\right)^\alpha \right],
\label{eq_shrinking}
\end{equation} 
where $q_0$, $0<q_0\le 1$ is a constant parameter that accounts for the probability of {\it leaving the ground level\/} and $\alpha>0$. 
To show  how the methodology must be deployed, we detail the simple case corresponding to $\alpha=1$, namely 
\begin{equation*}
q_{\ell}=\frac{q_0}{\ell+1}.
\end{equation*}
Figure~\ref{Fig_shrinking} shows how typical trajectories under this dynamics are less likely to access higher values than the initial model with $q_{\ell}=q_0$, cf. Fig.~\ref{Fig_sample}.

It follows that the reset probability satisfies  
\begin{equation}
\PP\left\{\tau^*=\tau\right\}= \frac{q_0^{\tau-1}}{\left(\tau-1\right)!} \left(1-\frac{q_0}{\tau}\right),\  1\leq \tau <\infty.
\end{equation}
This leads to a mean inter-reset time $\EE\left[\tau^*\right]=e^{q_0}$, and to the following renewal equation 
 for $p(\ell,\tau;0)$, cf. Eq.~\eqref{TP} with $\ell=0$, 
\begin{eqnarray}
p(\ell,\tau;0)&=&\frac{q_0^{\tau}}{\tau!} \delta_{\tau,\ell}\nonumber\\
&+&\sum_{k=1}^{\tau}\frac{q_0^{k-1}}{\left(k-1\right)!} \left(1-\frac{q_0}{k}\right) p(\ell,\tau-k;0).
\end{eqnarray}
Appropriate use of  the $z$-transform  permits solving this equation as 
\begin{equation}
p(\ell,\tau;0)=\frac{q_0^{\ell}}{\ell!} e_{\tau-\ell}(-q_0)\ \Theta(\tau- \ell).
\label{TP0SP_sol}
\end{equation}
Here we have introduced the exponential sum function $e_{n} (-q_0)$, defined  as the $n$-th Taylor polynomial for the exponential function:
\begin{equation}
e_{\tau} (-q_0) \equiv \sum_{k=0}^{\tau} \frac{(-q_0)^k}{k!}.
\end{equation}
Letting $\tau\to\infty$ in Eq.~\eqref{TP0SP_sol} we find that the stationary state 
has Poisson distribution with parameter $q_0$ and mean position $1/q_0$.  

The results can be generalized to an arbitrary election for $q_\ell$. Skipping the details, we note that for $\tau\geq \ell$
\begin{equation}
p(\ell,\tau;0)= \PP\left\{\tau^*>\ell\right\} \left[\delta_{\tau,\ell}+m(\tau-\ell)-m(\tau-\ell-1) \right]. 
\label{general1}
\end{equation} 
The renewal theorem implies then that as long as one has $\EE\left[\tau^*\right] <\infty$, the stationary distribution is  
\begin{equation}   
p(\ell)  = \frac{ \PP\left\{\tau^*>\ell\right\} } {\EE\left[\tau^*\right]},\  0\le \ell  <\infty. 
\label{general2}
\end{equation}  
Thus, both the finite-time position distribution and stationary state follow in closed form given the distribution of renewals $\PP\left\{\tau^*>\ell\right\}=q_0\cdots q_{\ell-1}$ and the function $m(\tau)$ |which can be recovered solving Eq.~\eqref{renew_m} appropriately generalized, cf. \cite{kt}.

\subsection{Sisyphus random walk on the integers}
Here we consider the Sisyphus random walk generalized to a case where trajectories could, {\it after every reset\/}, either increase or decrease linearly with probabilities $\rho$ and $1-\rho$, respectively.  Concretely, given $X_{t} = 0$ we  generalize Eq.~\eqref{process} to
 \begin{equation}  
 X_{t+1}=\left\{
\begin{array}{ll} 1,  &\text{with probability  }q\rho, \\-1, &\text{with probability  } q(1-\rho), \\ 0,  &\text{with probability  }1-q,  \end{array}\right.
\end{equation}
while if $X_{t}\ne 0$ then   
\begin{equation}
X_{t+1}=\left\{
\begin{array}{ll}
\ell+\text{sgn}(\ell), &\mbox{ with probability } q_{} ,\\
0,&\mbox{ with probability } (1-q_{}).
\end{array}
\right.
\label{Sisyphus random walk with two-sided reflection}
\end{equation}

$X_t$ is now  a {\it  Markov chain on the full integers\/}, namely $X_t\in\{-t,-t+1,\dots,t-1,t\}$ for $t~\in~\{0,1,2,\ldots\}$, defined by two independent parameters, $q$ and $\rho$, with $0<q<1$, $0<\rho< 1$. It  could be used as a crash model for search strategies wherein the walker may return to the origin and restart in the opposite direction.~\footnote{In this case, the rock pushed by the king of Ephyra rolls down toward a valley located between two twin hills.}  The optimal reset strategy is described below, see Eq.~\eqref{optimal_reset}. This more general situation may be analyzed with similar techniques to those employed previously. We find it  convenient to define 
\begin{equation}
\rho_\ell\equiv \left\{
\begin{array}{ll}  \rho, &\textrm{for } \ell>0,\\ 1, &\textrm{for } \ell=0,\\1-\rho,  &\textrm{for } \ell<0.\end{array}
\right.
\end{equation}
With such proviso we find that when $\ell_0=0$, Eq.~\eqref{TP} generalizes to
 \begin{equation}
p(\ell,\tau;0)=\rho_\ell q^{|\ell|} \delta_{\tau,|\ell|}+ \sum_{k=1}^{\tau}q^{k-1} (1-q)  p(\ell,\tau-k;0),
\end{equation} 
whose solution for  $-\tau\le \ell\le\tau$ reads
\begin{equation}
p(\ell,\tau;0)=  \rho_\ell   q^{|\ell|}\left[\delta_{\tau,|\ell|} + (1-q) \Theta\left(\tau-|\ell|-1\right)\right].
\end{equation}

Other statistical observables, like the propagator $p(\ell,\tau;\ell_0)$ or the first hitting time, can also be obtained by establishing the renewal equation that codifies the possible behaviors after the first renewal. Starting from $\ell_0\ne 0$, the propagator reads simply 
\begin{equation}
p(\ell,\tau;\ell_0)= q^{\tau} \delta_{\ell_0+ \tau  \sigma_0,\ell} +\rho_\ell q^{|\ell|}(1-q)  \Theta\left(\tau-|\ell|-1\right),
\end{equation}   
where $ \sigma_{0}\equiv\text{sgn} (\ell_0)$. The stationary state has a two-sided geometric distribution
\begin{equation}
p(\ell)  =\rho_\ell  (1-q)q^{|\ell|}, \  -\infty<\ell<\infty,
\end{equation}
and satisfies $ \PP\left\{X_{\tau^*-1}=\ell\right\}= \PP\left\{|X_\infty|= \ell\right\}$.

If $\ell > 0$, say,  Eq.~\eqref{FP_renew} for the hitting time reads now
 \begin{eqnarray}
\mathcal{P}(t,\ell) &=& \rho q^{ \ell} \delta_{t, \ell}+\sum_{k=1}^{\ell} q^{k-1} (1-q)  \mathcal{P}(t-k,\ell) \nonumber\\
&+&(1-\rho) \sum_{k=\ell+1}^{\infty}  q^{k-1} (1-q)  \mathcal{P}(t-k,\ell), 
\end{eqnarray}
with $t\ge \ell$, and hence the generating function and mean of the hitting time read respectively
\begin{equation}
\widehat{\mathcal{P}}(z,\ell)=\frac{\rho  \left(q z\right)^{\ell} (1-q z)}{1- z + \rho  z (1-q)   \left(q z\right)^{\ell} }, 
\end{equation}
and
\begin{equation}   \EE\left[\mathcal{F}(\ell)\right]= \frac{1}{ 1-q}\left(\frac{1}{\rho q^{\ell}}- 1\right).\label{mean_exit_two_sided}
\end{equation}

Equation~\eqref{mean_exit_two_sided} can be used to minimize the mean exit time~\cite{EM11a}, for given  $\rho$ and $\ell$. 
The condition for the minimum is
\begin{equation}   
\rho q^{\ell +1}- (1+\ell) q+\ell=0.
\label{optimal_reset}
\end{equation}
By a well-known result in calculus, this equation has exactly one solution $q^*$ satisfying $0\leq q^*\leq 1$. The election $q = q^*$ corresponds to the {\it optimal search, or reset, strategy\/} for the two-sided Sisyphus random walk. In particular, it reduces to
\begin{equation}   
\rho e^{\epsilon^*-1}=\epsilon^*,
\label{min_epsilon}
\end{equation}
when $\ell\to\infty$. In Eq.~\eqref{min_epsilon} we have introduced parameter $\epsilon^*$, $\epsilon^*\equiv(\ell+1) q^*-\ell$, $0<\epsilon^*<1$,
in terms of which the minimum mean exit time reads,
\begin{equation}   
\EE\left[\mathcal{F^*}(\ell)\right]\to \frac{\ell}{\epsilon^*}.
\end{equation}

\section{Conclusions}
\label{Sec_conclusions}

We have analyzed an extremely simple random walk with constant deterministic dynamics that may randomly return to the origin and have determined its main statistics; in particular, the first-passage time and high-water marks of the process are discussed. It turns out that this simple evolution law is misleading and  the corresponding dynamics is surprisingly complex and exhibits a rich behaviour.  Nevertheless suitable renewal ideas may be used to simplify the analysis. We suggest generalizations of the system, appropriate to different physical settings;  in particular a situation where the tendency to return increases with the distance to the initial state can be incorporated into our formalism. The resulting system is then capable to provide a gross description of Sisyphus cooling, and formation and growth of houses of cards and icicles. Depending on the election, Zipf, geometric or Poisson distributions are found to describe the equilibrium state.


\begin{acknowledgments}
The authors acknowledge support from the Spanish Ministerio de Econom\'{\i}a y Competitividad (MINECO) under Contract No. FIS2013-47532-C3-2-P (MM) and MTM2012-38445 (JV), and from the Ag\`encia de Gesti\'o d'Ajuts Universitaris i de Recerca (AGAUR), Contract No. 2014SGR608 (MM). We are also grateful to an anonymous referee whose comments helped to considerably improve and put the paper in perspective.
\end{acknowledgments}

\appendix

\section{Inverse $z$-transforms}
\label{zTransform}
We have
\begin{equation}
\widehat{\mathcal{P}}(z,\ell)\equiv\sum_{k=0}^{\infty}\mathcal{P}(k,\ell) z^k,
\end{equation}
with
\begin{eqnarray}
\widehat{\mathcal{P}}(z,\ell)&=&\left(q z\right)^{\ell}\frac{1-q z}{1- z+ (1-q)z\left(q z\right)^{\ell}}.
\end{eqnarray}
We can obtain $\mathcal{P}(k,\ell)$ by looking at the coefficient in front of the  $z^k$ term. Consider 
\begin{equation}
G(z)\equiv\frac{1}{1- z+ (1-q)z\left(q z\right)^{\ell}},
\end{equation}
in terms of which $\widehat{\mathcal{P}}(z,\ell)$ reads,
\begin{eqnarray*}
\widehat{\mathcal{P}}(z,\ell)=\left(q z\right)^{\ell}\left(1-q z\right) G(z).
\end{eqnarray*}
One has
\begin{eqnarray*}
G(z)&=&\frac{1}{1- z\left[1- (1-q)\left(q z\right)^{\ell}\right]}\\
&=&\sum_{m=0}^{\infty} \left[1-(1-q)\left(q z\right)^{\ell}\right]^m z^m\\
&=&\sum_{m=0}^{\infty} \sum_{k=0}^{m} \binom{m}{k}\left[(q-1)q^{\ell}\right]^k z^{m+k\ell}\\
&=&\sum_{k=0}^{\infty}\sum_{m=k}^{\infty} \binom{m}{k}\left[(q-1)q^{\ell}\right]^k z^{m+k\ell}. 
\end{eqnarray*}
Hence
\begin{eqnarray*}
\widehat{\mathcal{P}}(z,\ell)&=&q^{\ell} \sum_{k=0}^{\infty}\sum_{m=k}^{\infty} \binom{m}{k}\left[(q-1)q^{\ell}\right]^k z^{m+(k+1)\ell}\\
&-&q^{\ell+1} \sum_{k=0}^{\infty}\sum_{m=k}^{\infty} \binom{m}{k}\left[(q-1)q^{\ell}\right]^k z^{m+(k+1)\ell+1},
\end{eqnarray*}
and therefore, collecting the terms with $z^{\tau}$, we have
\begin{eqnarray*}
\mathcal{P}(\tau,\ell)&=&q^{\ell} \sum_{k=0}^{\left\lfloor\frac{\tau-\ell}{\ell+1}\right\rfloor} \binom{\tau-(k+1)\ell}{k}\left[(q-1)q^{\ell}\right]^k\\
&-&q^{\ell+1} \sum_{k=0}^{\left\lfloor\frac{\tau-\ell-1}{\ell+1}\right\rfloor}\binom{\tau-(k+1)\ell-1}{k}\left[(q-1)q^{\ell}\right]^k, 
\end{eqnarray*}
for $\tau \geq \ell+1$, 
and where $\lfloor x \rfloor=\max\, \{k\in\mathbb{Z}\mid k\le x\}$.
Then we have, for $\ell+1\leq \tau \leq 2\ell$, $\mathcal{P}(\tau,\ell)=(1-q)q^\ell$, for $2 \ell+1\leq \tau \leq 3\ell+1$,
\begin{equation*}
\mathcal{P}(\tau,\ell)=(1-q)q^\ell\left[1- (\tau-2\ell)(1-q)q^\ell-q^{\ell+1}\right],
\end{equation*}
for $3 \ell+2\leq \tau \leq 4\ell+2$,
\begin{eqnarray*}
\mathcal{P}(\tau,\ell)&=&(1-q)q^\ell\left[1- (\tau-2\ell)(1-q)q^\ell-q^{\ell+1}\right]\\
&+&\frac{\tau-3\ell-1}{2}(1-q)^2q^{3\ell}\left[(\tau-3\ell)(1-q)+2q\right],
\end{eqnarray*}
and so forth. For $\widehat{\mathcal{P}}(z,\ell;\ell_0)$, we have
\begin{eqnarray*}
\widehat{\mathcal{P}}(z,\ell;\ell_0)&=&\frac{(1-q) z \left(q z\right)^{\ell}}{1- z+ (1-q)z\left(q z\right)^{\ell}}\\
&=&(1-q) z \left(q z\right)^{\ell}G(z),
\end{eqnarray*}
for $\ell\le \ell_0$, and
\begin{eqnarray*}
\widehat{\mathcal{P}}(z,\ell;\ell_0)&=&\frac{(1-z)\left(q z\right)^{\ell-\ell_0}+(1-q) z \left(q z\right)^{\ell}}{1- z+ (1-q)z\left(q z\right)^{\ell}}\\
&=&\left[(1-z)\left(q z\right)^{\ell-\ell_0}+(1-q) z \left(q z\right)^{\ell}\right]G(z),
\end{eqnarray*}
for $\ell>\ell_0$; therefore we can use the same technique. In particular, for the case in which the two records are consecutive, $\mathcal{R}(\tau,\ell)=\mathcal{P}(\tau,\ell;\ell-1)$,  $\tau\geq 1$, we have
\begin{eqnarray*}
\mathcal{R}(\tau,\ell)&=&q \sum_{k=0}^{\left\lfloor\frac{\tau-1}{\ell+1}\right\rfloor} \binom{\tau-k\ell-1}{k}\left[(q-1)q^{\ell}\right]^k\\
&-&\Theta(\tau-2)q\sum_{k=0}^{\left\lfloor\frac{\tau-2}{\ell+1}\right\rfloor}\binom{\tau-k\ell-2}{k}\left[(q-1)q^{\ell}\right]^k\\
&+&\Theta(\tau-\ell-1)(1-q)q^{\ell}\\
&\times&\sum_{k=0}^{\left\lfloor\frac{\tau-\ell-1}{\ell+1}\right\rfloor}\binom{\tau-(k+1)\ell-1}{k}\left[(q-1)q^{\ell}\right]^k.
\end{eqnarray*}

\end{document}